\documentclass[preprint, endfloats ,showpacs,amsmath,amssymb,apl,aps,superscriptaddress, floatfix]{revtex4}
\pdfoutput=1
\usepackage{color}
\usepackage{graphicx}
\usepackage{xspace}
\usepackage{multirow}
\newcommand{\x}{\,$\times$\,}

\begin{document}

\setlength{\topmargin}{-1cm}

\title{Chiral symmetry breaking of magnetic vortices by sample roughness}

\author{A. Vansteenkiste}
\affiliation{
  Department of Subatomic and Radiation Physics, Ghent University, Proeftuinstraat 86, 9000 Gent, Belgium.
}

\author{M. Weigand}
\affiliation{
  Max Planck Institute for Metals Research, Heisenbergstr. 3, 70596 Stuttgart, Germany.
}

\author{M. Curcic}
\affiliation{
  Max Planck Institute for Metals Research, Heisenbergstr. 3, 70596 Stuttgart, Germany.
}

\author{H. Stoll}
\affiliation{
  Max Planck Institute for Metals Research, Heisenbergstr. 3, 70596 Stuttgart, Germany.
}

\author{G. Sch{\"u}tz}
\affiliation{
  Max Planck Institute for Metals Research, Heisenbergstr. 3, 70596 Stuttgart, Germany.
}

\author{B. {Van Waeyenberge}}
\affiliation{
  Max Planck Institute for Metals Research, Heisenbergstr. 3, 70596 Stuttgart, Germany.
}

\date{\today}

\begin{abstract}

Finite-element micromagnetic simulations are employed to study the chiral symmetry breaking of magnetic vortices, caused by the surface roughness of thin-film magnetic structures.  An asymmetry between vortices with different core polarizations has been experimentally observed for square-shaped platelets. E.g., the threshold fields for vortex core switching were found to differ for core up and down. This asymmetry was however not expected for these symmetrically-shaped structures, where both core polarizations should behave symmetrically. Three-dimensional finite element simulations are employed to show that a small surface roughness can break the symmetry between vortex cores pointing up and down. A relatively small sample roughness is found sufficient to reproduce the experimentally observed asymmetries. It arises from the lack of mirror-symmetry of the rough thin-film structures, which causes vortices with different handedness to exhibit asymmetric dynamics.

\end{abstract}


\maketitle

\section{Introduction}

A magnetic vortex is a typical ground state of soft magnetic thin-film platelets with dimensions in the micron or submicron range and, e.g., a square or circular shape. The magnetostatic field tries to align the magnetization parallel to the sample boundaries, so it is forced into the sample plane and curls around the structure-center, forming the vortex. At the center of the vortex, however, the exchange interaction causes the magnetization to turn out of the sample plane, avoiding an antiparallel alignment of neighboring moments.  \cite{feldtkeller65, raabe00}.\\

The in-plane curling magnetization can have two possible senses of circulation $c$: clockwise or counterclockwise. These states are denoted by $c=-1$ and $c=+1$, respectively. Analogously, the out-of-plane magnetization of the vortex core can have two possible polarizations $p$: either up or down, denoted by $p=+1$ and $p=-1$, respectively. The combination of $c$ and $p$ also gives the vortex a three-dimensional handedness. A vortex with $c\times p=+1$ is defined to be right-handed, and one with $c\times p=-1$ to be left-handed \cite{choe04}.\\


Magnetic vortices are interesting configurations from a fundamental point of view, but are also possible candidates for future memory applications. The core polarization $p$, which can store one bit of information, can  be switched by low-amplitude magnetic fields \cite{vanwaeyenberge06}. In-plane alternating \cite{vanwaeyenberge06} or rotating \cite{curcic08} magnetic fields can switch the core easily when they resonantly excite the so-called vortex gyrotropic mode. This is the mode that corresponds to a circular movement of the vortex core around the structure center \cite{huber82, argyle84}. It typically has a resonance frequency in the sub-GHz range and a width in the order of tens of MHz \cite{novosad05}.\\

In a square or disk-shaped sample, all combinations of $c$ and $p$ are symmetric, i.e., these states have the same energy  and symmetric dynamics. In particular their gyrotropic eigenfrequency and switching thresholds are exactly equal \cite{guslienko02}. This symmetry follows from the mirror-symmetry of the circular or square sample shape. E.g., the $p=+1$ and $p=-1$ states can be transformed into each other by mirroring the sample along the sample plane (defined to be the $xy$-plane, $z$ is the out-of plane direction). Similarly the $c=+1$ and $c=-1$ states can be transformed into each other by mirroring along the $xz$ or $yz$ plane.\\

However, several experiments on Permalloy (Ni$_{80}$Fe$_{20}$) square platelets have revealed a relatively large asymmetry in the dynamics of a vortex core when its polarization was inverted. First of all, during a steady-state gyration of the vortex core, induced by an in-plane alternating field, it was found that the gyration radius was significantly different for the two core polarizations \cite{chou07}. Secondly, the required amplitudes for switching with rotating fields were found to differ for the two polarizations: the switching thresholds of the investigated sample differed by more than 10\% for $p=\pm1$  \cite{curcic08}.\\

Since the investigated structures were geometrically symmetric, a possible explanation for the $p$-symmetry-breaking was sample roughness, which breaks the mirror-symmetry on a small length scale. This was suggested as a possible explanation in the abovementioned papers \cite{chou07, curcic08}, although it remained unclear if a small roughness could cause such a large $p$-asymmetry.\\

Surface roughness is usually inevitable in evaporated or sputter-deposited thin films, so its influence on the vortex dynamics may have to be taken into account for the development of, e.g., vortex-based memories \cite{kim08}. Especially the different vortex core switching thresholds for core up/down \cite{curcic08} may have implications for such devices.\\

\section{Micromagnetic Simulation Setup}

In this work, a micromagnetic study of symmetry breaking in vortices by a chiral sample geometry is presented, with surface roughness as a particular case. An in-house developed 3D finite-element program \footnote{\textsc{amumag} micromagnetic package, code available at {http://code.google.com/p/amumag}} was employed  to solve the Landau-Lifshitz equation \cite{landau35}. The finite elements allow the mesh to be irregular, so that arbitrary sample shapes can be well approximated. In order to efficiently calculate the magnetostatic field on such a mesh, the Fast Multipole Method \cite{visscher04} was used. Special care was taken so that the small truncation errors of this method would not break the mirror-symmetry of the sample. Material parameters typical for Permalloy were chosen: exchange constant $A$=13\x 10$^{-12}$\,J/m, saturation magnetization $M_s$=736\x10$^3$\,A/m \cite{vansteenkiste08}, damping parameter $\alpha$=0.01, anisotropy constant $K_1$=0, gyromagnetic ratio $\gamma$=2.211\x10$^5$\,m/A\,s. A 3D discretization was performed with a cell size of approximately 3.9\,nm\,$\times$\,3.9\,nm in x and y, and four layers in z, corresponding on average to 12.5\,nm (see Fig. \ref{rough}(c)). \\

\section{Simulation of chiral geometries}

\begin{figure}[!htb]
\centering
\includegraphics[width=1\linewidth]{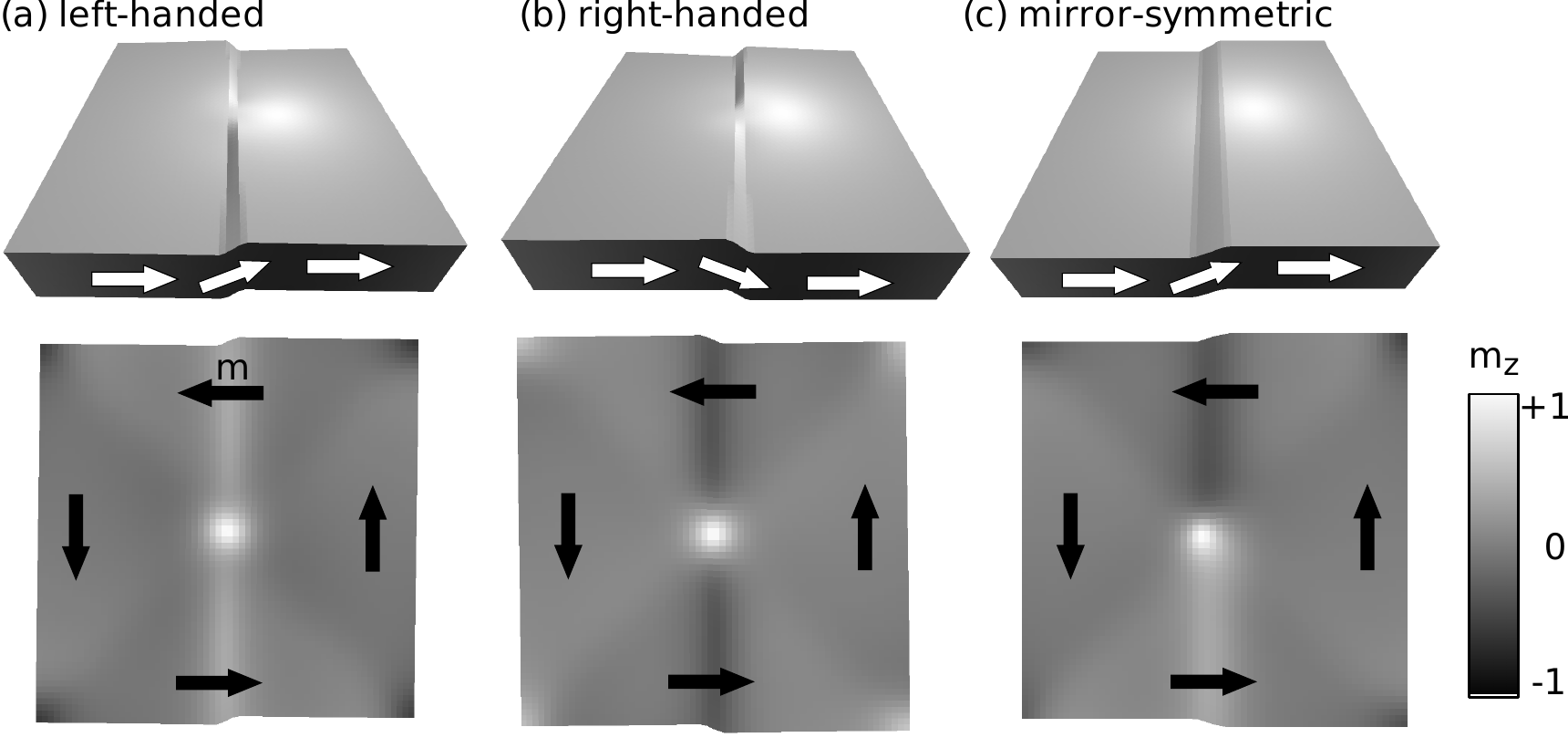}
\caption{\label{helix} \textbf{Row 1:} deformed thin-film platelets. The structures (a) and (b) are chiral and are defined to be left-handed and right-handed, respectively, while sample (c) is mirror-symmetric and thus has no handedness. \textbf{Row 2:} out-of-plane component $m_z$ of the magnetization in those structures, for $c=+1$, $p=+1$. In the chiral structures (a) and (b), $m_z$ at the position of the jumps points either twice up or twice down, so no symmetry exists between the magnetization of (a) and (b). In the mirror-symmetric structure (c), $m_z$ points once up and once down, so that here a symmetry between the two possible $p$-states remains.}
\end{figure}

\newcommand{\highlight}[1]{\textbf{#1}}
\begin{table}

sample (a): left-handed geometry

\begin{tabular}{|l||l|l|}
\hline
$cp$ & 1 & -1 \\\hline\hline
$f_0$ & 914.9\,MHz & 837.8\,MHz \\\hline
$E_0$ & 2.97$\times$10$^{-17}$\,J & 2.99$\times$10$^{-17}$\,J\\\hline
\end{tabular}

sample (b): right-handed geometry

\begin{tabular}{|l||l|l|}
\hline
$cp$ & 1 & -1 \\\hline\hline
$f_0$ & 837.8\,MHz & 914.9\,MHz \\\hline
$E_0$ & 2.99$\times$10$^{-17}$\,J & 2.97$\times$10$^{-17}$\,J \\\hline
\end{tabular}

sample (c): mirror-symmetric

\begin{tabular}{|l||l|l|}
\hline
$cp$ & 1 & -1 \\\hline\hline
$f_0$ & 1054.4\,MHz & 1054.4\,MHz \\\hline
$E_0$ & 2.94$\times$10$^{-17}$\,J & 2.94$\times$10$^{-17}$\,J \\\hline
\end{tabular}

\caption{\label{tab2} Simulated resonance frequencies $f_0$ of the vortex gyrotropic mode and ground state energies $E_0$, for the 250\,nm\,$\times$\,250\,nm\,$\times$\,50\,nm structures shown in Fig. \ref{helix}. The four combinations of vortex circulation $c$ and polarization $p$ were simulated; states with equal $c \times p$ yielded the same results. Structures (a) and (b) are chiral and are each other's mirror image, while structure (c) is mirror-symmetric (see Fig. \ref{helix}). An asymmetry between vortices with a different handedness $c \times p$ arises only for the chiral structures.}
\end{table}

To understand the properties of magnetic vortices in chiral nanostructures, a first series of simulations has been carried out on three deformed platelet geometries, shown in row 1 of Fig. \ref{helix}. These were 250\,nm $\times$ 250\,nm wide and 50\,nm thick. Sample (a) is a screw-like structure which is not mirror-symmetric, and we arbitrarily define its chirality to be left-handed. Sample (b) is obtained by mirroring sample (a) and is therefore defined as a right-handed structure. Sample (c), on the other hand, also has two ``jumps'', similar to (a) and (b), but this sample is mirror-symmetric.\\

The gyrotropic eigenfrequencies and ground-state energies of these samples were simulated for the four possible combinations of $c$ and $p$, and are listed in Tab. \ref{tab2}. In the mirror-symmetric sample (c) all combinations of $c$ and $p$ yielded the same properties, while a $c$- and $p$-asymmetry was found for the chiral samples (a) and (b). In those cases, the eigenfrequency and ground-state energy changed when the vortex core polarization $p$ was inverted.  The exact same asymmetry was observed between states with different circulations $c$, and the asymmetry vanishesd again when both $c$ and $p$ were inverted at the same time. This residual $cp$-symmetry is fundamental and follows immediately from micromagnetic theory: inverting both $c$ and $p$ corresponds to inverting the magnetization $\vec{m}$, which in the absence of an external field does not change the energetics of a magnet. The $c$ and $p$ asymmetries can thus be interpreted in terms of chirality: vortices with equal handedness $c \times p$ are fundamentally symmetric, but in a chiral sample geometry, vortices with opposite handedness are asymmetric.\\

Note that mirroring the handedness of the vortex has the same effect on its properties as mirroring the geometry of the sample. E.g., a left-handed vortex in a left-handed sample (a) has the same properties as a right-handed vortex in a right-handed sample (b). When we denote a left-handed structure by $h=+1$, a right-handed structure by $h=-1$ and a mirror-symmetric structure by $h=0$, then all configurations with equal $c \times p \times h$ are symmetric. In particular, mirror-symmetric structures are always symmetric for all combinations of $c$ and $p$, and in all cases the fundamental $cp$-symmetry exists: vortices with equal handedness are always symmetric.\\

In the absence of a magnetic field, a chiral sample geometry was thus found to be necessary for breaking the $c$ and $p$-symmetries. This necessary condition can be intuitively understood by investigating out-of-plane component $m_z$ of the magnetization. This is shown in Fig. \ref{helix}, for the three respective sample geometries. Due to the shape anisotropy, which wants to align the magnetization parallel to the sample boundaries, the magnetization acquires an out-of-plane component $m_z\neq 0$ in the regions where the films are deformed. These regions couple to the vortex core via the exchange interaction and the dipole-dipole coupling. As a consequence, vortex cores with different polarizations can have different energies in such an environment. When either the in-plane circulation $c$, or the core polarization $p$ is inverted, the coupling changes sign as well. The asymmetry can thus be understood as a $cp$-coupling due the chiral sample geometry.\\

Row 2 of Fig. \ref{helix} shows that in sample (a), the left-handed circulation causes the extra out-of-plane magnetization at each of the jumps to point up. The coupling between those regions and the vortex core happens to be favorable in this case, causing this configuration to have the lowest energy. In the right-handed structure (b), the opposite takes place. The out of-plane magnetization at the jumps is antiparallel to the vortex core, causing the configuration to have a higher energy. On the other hand, if the sample is mirror-symmetric like sample (c), The magnetization points up at one jump and down at the other one, so that the coupling is equal for core up and down.\\

Note that the presence of an external field in the sample plane can cause an additional symmetry breaking. E.g., in the mirror-symmetric sample (c), a static field in the $x$-direction will displace the vortex core down, towards the region with a positive out-of plane magnetization. It can easily be understood that an up vortex will behave differently from a down vortex in such an environment. In this particular case, the energy of the up vortex was found to be lower. However, even in the presence of an in-plane magnetic field, the deformed geometry is still necessary to break the symmetry between the four vortex states.\\

\section{Simulation of roughness}

The previous set of simulations was artificially chosen to illustrate the mechanism behind the $c$ and $p$ symmetry-breaking by a lack of geometric mirror-symmetry. Random surface roughness is a particular case where the mirror-symmetry is (almost always) broken, and where the condition for a $cp$-asymmetry is thus fullfilled. How large this asymmetry is and which vortex handedness is favored, can be expected to depend the details of the roughness.\\

\begin{figure}[!htb]
\centering
\includegraphics[width=1\linewidth]{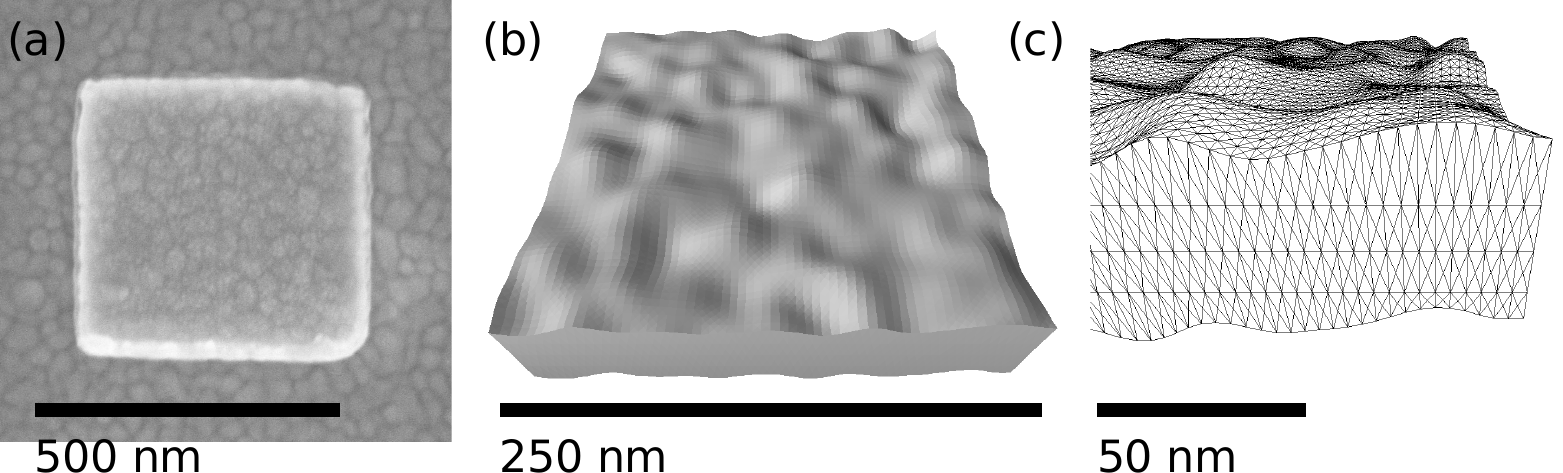}
\caption{\label{rough} (a) Scanning electron micrograph of a typical 500\,nm Permalloy platelet in which a $p$-asymmetry was observed. The platelet lies on top of a rough copper stripline. (b) Simulation geometry of a 250\,nm platelet with a roughness which mimics the effect of a rough underground. (c) Detail of (b), showing the 3D finite-element discretization.}
\end{figure}

To illustrate that a chiral sample geometry can explain the experimentally observed asymmetries, we performed a second set of calculations of samples with a random surface roughness, similar to the samples investigated in the abovementioned experiments \cite{chou07, curcic08}. The simulated geometry was inspired on a scanning electron micrograph of a typical Permalloy sample investigated in those experiments, shown in Fig. \ref{rough}(a). The sample lies on top of a relatively rough copper stripline, which is used for the magnetic excitation \cite{vanwaeyenberge06, chou07, curcic08}. Therefore, a square platelet was simulated as if it were deposited on a rough underground while conserving its thickness. A peak amplitude of 10\,nm was chosen for this roughness. To make the geometry more realistic, a smaller additional roughness of 5\,nm was added to the top surface, although this did not significantly change the outcome of the simulations. The length scale of the roughness was also chosen to mimic the electron micrograph in Fig. \ref{rough}.\\


\begin{figure}[!htb]
\centering
\includegraphics[width=0.7\linewidth]{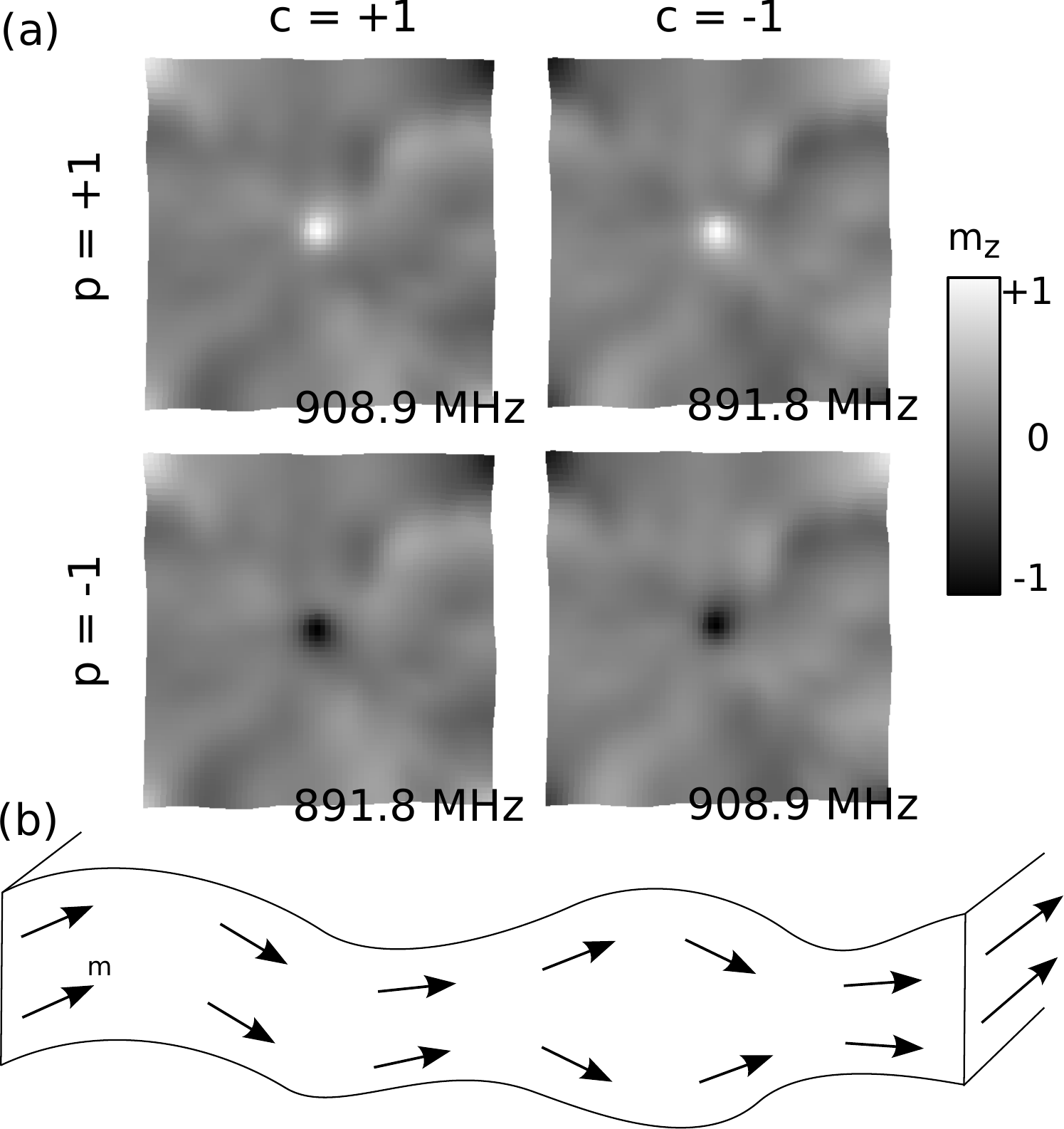}
\caption{\label{coupl} \textbf{(a)} Out-of-plane component $m_z$ of the magnetization in a rough sample, for the possible combinations of $c$ and $p$. Additional out-of-plane moments arise, which are not present in perfectly flat structures. Their direction is dependend on the circulation $c$. These out-of-plane moments interact with the vortex core via the exchange and dipole-dipole coupling, resulting in different energies for a vortex that points up or down. \textbf{(b)} Illustration how these out-of-plane moments arise. The shape anisotropy keeps the magnetization parallel to the sample edges, locally tilting the magnetization up or down at some positions. The resulting out-of-plane moments are especially large when the top and bottom surface are nearly parallel, as is caused by a rough underlayer.}
\end{figure}

Due to the random nature of the roughness, the effect in different samples can be expected to have a different strength or might incidentally even vanish. It lies however not in the scope of this work to study the differences between many rough geometries, neither to exactly reproduce the detailed roughness of the experimentally investigated structures. We only want to illustrate that realistic roughness parameters are sufficient to explain the experimentally observed asymmetries. Therefore, we have simulated three rough geometries based on different random number seeds, and we present the simulation with the median symmetry breaking between the gyrotropic eigenfrequencies in Fig. \ref{coupl}. The expected asymmetry between the states with different vortex handednes $c \times p$ was found. The figure also shows the out-of plane moments that arise due to the roughness. A side view of a vortex structure in a rough thin film (Fig. \ref{coupl} (a)) illustrates how these out-of-plane moments are especially pronounced when the full layer happens to be bent upwards or downwards, which is exactly the case when the film is deposited on a rough underground. The expected asymmetry was also found in the ground-state energy: for $c \times p = +1$ this was 3.0151$\times$10$^{-17}$\,J, while for $c \times p = -1$ it was 3.0102$\times$10$^{-17}$\,J.\\

The observed difference in resonance frequencies of about 20\,MHz is in the order of the width of the resonance, and is thus expected to significantly affect the dynamics.  E.g., when the excitation is chosen in resonance for the up vortex, it may already be well out of resonance for the down vortex. The small asymmetry in gyration frequency could thus cause a relatively large asymmetry in the response to an alternating field near the resonance frequency.\\

This has been quantified by applying a rotating magnetic field in gyrotropic sense of the vortex core \cite{curcic08} to each of the $c,p$ states. The field was slowly ramped up with a time constant of 15\,ns, until it was strong enough to switch the vortex core polarization. A frequency of 908\,MHz was chosen, which lies closest to the eigenfrequency of the $cp=+1$ configurations. For the different combinations of $c$ and $p$, switching thresholds between 4.19\,mT and 4.69\,mT were found. This asymmetry between the switching thresholds is of the same order of magnitude as the experimentally observed asymmetry of about 10\% \cite{curcic08}. The presence of a small roughness can thus explain the experimental asymmetries. Note that in the absence of roughness, all combinations of $c$ and $p$ yielded exactly the same switching threshold and gyrotropic eigenfrequency.\\

Finally, it was asserted that the sample size has no critical effect on the symmetry breaking, by simulating three 500\,nm\,$\times$\,500\,nm$\times$\,40\,nm squares with random roughness. Due to computational limitations, these could only be discretized in two dimensions, but an asymmetry of the same order of magnitude was found. This reflects the local nature of the effect of the roughnss on the vortex core.\\

\section{Conclusion}

We have presented micromagnetic simulations showing that the absence of mirror-symmetry in the sample shape can break the symmetry between vortices with opposite handedness $c\times p$. More specifically, inverting either the in-plane circulation $c$, the core polarization $p$ or the chirality of the sample shape changes the vortex energy and dynamics in an asymmetric way. A particular case of chiral sample geometry is a random surface roughness. The out-of-plane moments caused by the roughness couple with the vortex core, breaking the symmetry between core up and down. Although the global symmetry breaking by the roughness is small, the coupling of the vortex core with the local out-of-plane moments is strong enough to explain the experimentally observed asymmetries. Although we have shown that roughness \emph{can} break the chiral symmetry of vortices, this work does not exclude the possibility of other, additional sources of symmetry breaking. 

\section{Acknowledgments}
Financial support by The Institute for the promotion of Innovation by Science and
Technology in Flanders (IWT-Flanders) and by the Research Foundation Flanders (FWO-Flanders)
through the research grant 60170.06 are gratefully acknowledged.

\bibliography{biblio}

\end{document}